\documentclass[conference,twoside,10pt]{IEEEtran}
\usepackage{indentfirst}
\usepackage{amsmath}
\usepackage{amsfonts,color,xcolor}
\usepackage{cite}
\usepackage{times}
\usepackage{dsfont}
\usepackage{multirow}
\usepackage[pdftex]{graphicx}
\usepackage[caption=false]{subfig}
\IEEEoverridecommandlockouts

\newcommand{\bi}{\begin{itemize}}
\newcommand{\ei}{\end{itemize}}

\newcommand{\be}{\begin{eqnarray}}
\newcommand{\ee}{\end{eqnarray}}

\newcommand{\figw}{0.92\columnwidth}
\newcommand{\figwb}{0.9\columnwidth}
\newcommand{\figwa}{0.245\textwidth}

\newcommand{\deriva}{\partial}

\let\OLDthebibliography\thebibliography
  \renewcommand\thebibliography[1]{
  \OLDthebibliography{#1}
  \enlargethispage{2mm}
  \vspace{-0.8mm}
  \setlength{\parskip}{0pt}
  \setlength{\itemsep}{-0.2pt}
}

\begin{document}

\title{The Role of Feedback in AoI Optimization \\Under Limited Transmission Opportunities}
\author{
\IEEEauthorblockN{Andrea Munari}
\IEEEauthorblockA{Institute of Communications and Navigation\\
German Aerospace Center (DLR), We\ss ling, Germany \\
email: andrea.munari@dlr.de}
\and
\IEEEauthorblockN{Leonardo Badia}
\IEEEauthorblockA{Dept. of Information Engineering\\
University of Padova, Italy \\
email: leonardo.badia@unipd.it}
\thanks{The work was in part supported by the Federal Ministry of Education and Research of Germany in the programme of ``Souver\"{a}n. Digital. Vernetzt.'' Joint project 6G-RIC, project identification number: 16KISK022.}
}
\date{}
\maketitle
\thispagestyle{empty}
 \pagestyle{empty}

\begin{abstract}
Scheduling updates from remote sensors is a key task for the internet of things (IoT). In this context, the mathematical concept of age of information is often used to capture
the freshness of received data. This is, in turn, relevant to optimize the frequency
of the exchanges, especially for resource constrained (e.g., energy-limited) sensors.
Most investigations on the subject assume that the transmitter can leverage knowledge of the age of information at the receiver side to decide when to send data,
even when the communication channel is unreliable.
In reality, tracking the outcome of the updates would require additional consumption of resources  to acquire a feedback.
We investigate the optimal schedule of updates over a finite time horizon for a resource-constrained sensor that is allowed to perform a limited number of updates, as typical of IoT devices. We discuss the role of the feedback from the receiver, and whether it is convenient to ask for it whenever this causes
additional energy consumption and consequently allows the transmission of a lower number of updates. We analytically identify regions for the feedback cost and the reliability of the channel where making use of feedback may or may not be beneficial.
\end{abstract}

\begin{IEEEkeywords}
Age of Information; Internet of Things; Data acquisition; Feedback; Sensor networks.
\end{IEEEkeywords}

\section{Introduction}
\label{introduction}

A widely studied problem in Internet of things (IoT) systems in the last few years concerns the scheduling of status updates over
a communication channel between a transmitter, usually a remote sensor, and a receiver. The goal is typically to minimize the so-called age of information (AoI) metric at the receiver 
\cite{yates2021age,kaul2012real,kadota2016minimizing},
capturing how fresh or up-to-date the available perception of the monitored process is.

Focusing on AoI as the performance indicator enables a characterization more suitable for remote sensing applications than what achieved by standard metrics such as throughput or delay, while at the same time retaining the advantages and the beauty of a precise mathematical description
\cite{zhou2019minimum,zhou2022performance,kam2018age,sun2019closed}.
Usually, the long-term average AoI is specifically considered, even though there are variations
such as the peak AoI, discounted AoI, or other similar metrics \cite{badia2022discounted,mankar2021spatial,yavascan2021analysis}. Also, the many studies in the field may have subtle differences
in considering a discrete or a continuous time-axis, with subsequent modifications on the mathematical
tools employed such as Markov chains or renewal theory \cite{ceran2019average,hatami2021aoi,feng2021age,tang2020minimizing,B}.

In this paper, we discuss the average AoI minimization over a finite time horizon for a resource-constrained
transmitter-receiver system exchanging data over an error-prone channel. In this setting, we tackle the case in which only the receiver, and not necessarily the transmitter, knows about the outcome of the communication
and whether it succeeded in bringing a useful update that lowered the AoI.
These specific variations are actually relevant in practical IoT scenarios,
especially considering modern commercial technologies such as LoRaWAN \cite{lora}, where transmissions must be scarce and listening for a feedback message can be as expensive as transmitting data.

In more detail, we consider a finite-time horizon, which corresponds to the
desired duration of a monitoring task. This departure from the commonly employed infinite-horizon modelling approach is, in our opinion, more consistent with the way of operation of practical IoT devices, which are assigned tasks over a finite time span.
Within this interval, the transmitter is allowed to send a limited (usually small) number of updates, mostly due to hardware and cost constraints.\footnote{Many IoT technologies such as LoRa have technological and/or legal limitations that prevent them from going beyond strictly bounded, low duty-cycles (e.g., in the order of $1\%$ for operations in the ISM band) \cite{lora}.}

For this scenario, we present two different analytical formulations depending on the availability at the transmitter's side of feedback about the success of updates. We consider first an offline (stateless) optimization that
schedules the instants for transmitting an update so as to minimize the expected average AoI across the finite horizon; this is done without exploiting the online information about the actual AoI, which is not available. Afterwards, we consider a stateful optimization of the scheduling instants with an online approach, based on dynamic programming, assuming the sender to receive feedback on the outcome
of each performed transmission.

Observing that an implementation of feedback may actually decrease the number of available transmission opportunities over the finite time horizon, e.g., due to additional energy consumption \cite{xie2020age,gindullina2021age}, we tackle the fundamental question of whether leaning on a return channel is truly beneficial in terms of AoI over a statless optimization.

To this aim, we provide an analytical characterization of the regions where using the feedback is advantageous (or not), depending on the time horizon, the number of opportunities, the transmission success probability, and the feedback cost.
These results can be of significant utility in practical contexts when setting up the operation policies for fresh status update from resource-constrained IoT devices \cite{zhou2022performance,badia2022discounted}, potentially establishing feedback policies that may be actually unnecessary.

The rest of this paper is organized as follows. In Section \ref{sec:rel}, we review related works, while Section \ref{sec:mod} illustrates our model and approach, divided into the two scenarios, without and with feedback.
Section \ref{sec:res} reports and discusses some numerical results to highlight the fundamental trade-offs of the setting, and finally Section \ref{sec:concs} draws  conclusions and points at possible extensions of the work.

\section{Related Works}
\label{sec:rel}
In the recent literature, many papers investigate the AoI in communication systems, especially in the context of remote sensing for the IoT
\cite{kaul2012real,kadota2016minimizing,zhou2019minimum,zhou2022performance,sun2019closed,kam2018age,mankar2021spatial,yavascan2021analysis,ceran2019average,hatami2021aoi,xie2020age,fountoulakis2020optimal,liu2021aion,li2020aoi,badia2021munari,tang2020minimizing,gindullina2021age,feng2021age,munari,badia2022discounted,pan2020timely}.
Taking AoI as a performance metric may lead to various formal approaches, relating to different degrees with the present paper.

A first class of significant works analyze the theoretical evaluation of AoI for different medium access policies and/or queueing disciplines \cite{kaul2012real,mankar2021spatial,badia2021munari}. This is possibly the most fundamental investigation, which is somehow orthogonal to our analysis, where we just consider a single node sending updates and we focus on the role of the feedback that enables the transmitter to make decisions based on the AoI at the receiver side. While in our model all the uncertainty about the transmission process is assumed, for the sake of simplicity, to be the result of Bernoulli-distributed independent errors, the same considerations can be extended to comprise other factors involving the presence of multiple nodes and realistic communication protocols \cite{munari,hatami2021aoi,zhou2019minimum}.

A second group of contributions consider, on the other hand, the problem of scheduling updates from a remote sensor accounting for AoI. For example, in \cite{kadota2016minimizing,li2020aoi,yavascan2021analysis,kam2018age,fountoulakis2020optimal,liu2021aion,pan2020timely} AoI is included as a threshold constraint to be respected by the scheduling, with different formalizations of the resulting optimization problems, solved either from a linear programming perspective or also advancing practical policies based on greedy or consecutive scheduling.

Another common setup involves the inclusion of the AoI within the objective function, usually to minimize the long-term average AoI at a sink node under constraints on the average number of transmissions at the source node \cite{ceran2019average,tang2020minimizing}, or related to energy harvesting \cite{hatami2021aoi,feng2021age,xie2020age,gindullina2021age}.

In particular, \cite{ceran2019average} compares automatic retransmission request (ARQ) policies. The use of feedback is implicitly considered, as required by standard or hybrid ARQ. However, the focus of \cite{ceran2019average} is on minimizing the long-term average AoI with an infinite horizon, and the specific scheduling pattern over a finite window is not considered. Moreover, costs of feedback are not directly addressed, focusing rather on the advantages of retransmission-based techniques. Our analysis differentiates itself by explicitly tackling the practical constraint of the price undergone to trigger the benefits of feedback.

In \cite{hatami2021aoi}, an approach based on dynamic programming and Markov decision processes can be found, akin to our study. 
Another similar analysis is in \cite{feng2021age}, where online optimal policies with and without feedback are compared, similar to what we do here, and the authors also derive a conclusion of optimality for a uniformly spread update policy. In both cases, the transmission constraints that limit the updates relate to energy harvesting, and an infinite time horizon is considered.

The focus of \cite{xie2020age} is on computing age and energy consumption and track their evolution for a coding and ARQ technique suitable to IoT scenarios, as opposed to considering a minimization problem. In \cite{gindullina2021age}, the problem is expanded, with the additional dimension of choosing from different energy sources.
AoI-optimal scheduling is also considered in \cite{zhou2019minimum,sun2019closed,tang2020minimizing}, specifically addressing IoT scenarios where multiple nodes with massive access are considered, with the focus being on different packet sizes, power availability, or random access techniques, respectively. Yet, the objective is once again the minimum long-term average AoI.

We remark that, from the perspective of dynamic programming, all stateful optimizations with the objective of infinite-term average AoI minimization often boil down to a threshold policy, i.e., to transmit when the AoI is overtaking a given value (and, if energy is considered, when the battery level is high enough). For our scenario with a finite horizon, the optimal scheduling policy changes over time, since transmission opportunities must be used sparingly at the beginning, but it is convenient to use all of them by the end of the time window. This marks a consequence of our analysis not highlighted by the aforementioned papers, i.e., for a finite time horizon, the knowledge of the time instant (which does not itself require a feedback) may be enough, thereby making the feedback superfluous even if the channel is unreliable.

Other notable related approaches involve using the concept of age of incorrect information (AoII) as in \cite{maatouk2020age} or tracking the dynamic evolution of AoI when the channel is inherently unreliable \cite{zhou2022performance}.  These contributions implicitly acknowledge for the cost of obtaining a feedback, which is reflected in the final measured metric rather than a price to pay in advance, which is the approach that we use here. An interesting development could be to link these kinds of investigations whenever the choice for receiving and exploiting a feedback is made available in the application design.

\section{System Model and Analysis}
\label{sec:mod}
We consider a transmitter/receiver pair interacting over an error-prone communication channel. The transmitter (also called source or sensor) sends data to the receiver (alternatively referred to as the destination or sink), reporting about the status of a monitored process, with the ultimate goal of keeping the information available at the sink as up-to-date as possible \cite{ceran2019average}.

\begin{figure}
  \centering
  \includegraphics[width=\figwb]{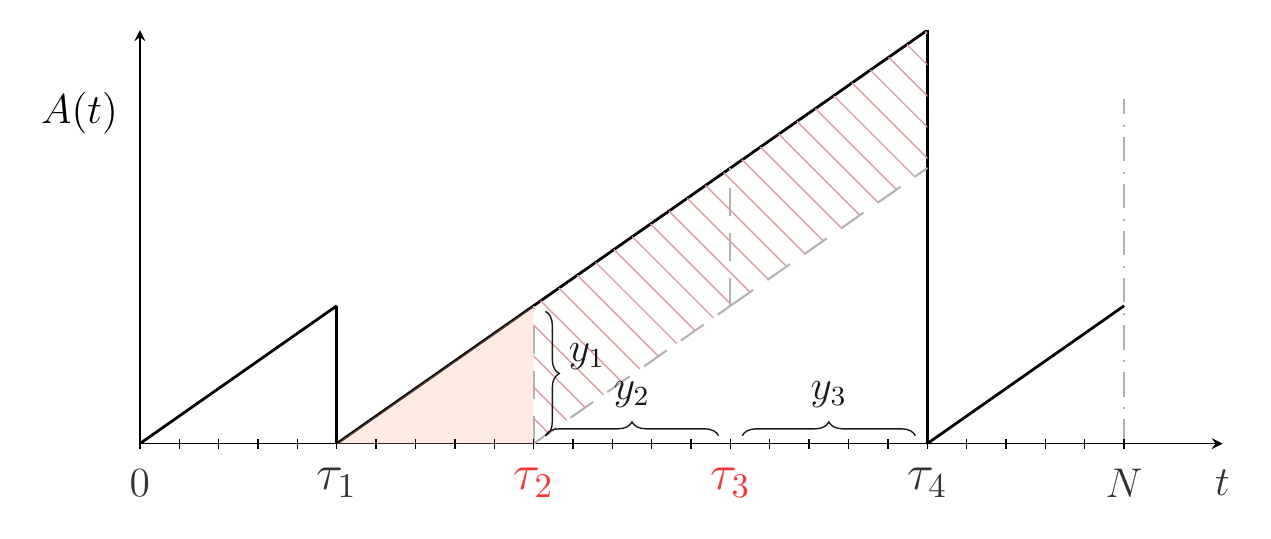}
  \vspace{-0.7em}
  \caption{Example timeline for the AoI evolution over time. In the considered case, four transmissions are performed over a time horizon of duration $N$, and the second and third transmissions fail.}
  \label{fig:timeline_aoi}
    \vspace{-0.5em}
\end{figure}

For convenience, we divide time into epochs, and assume that an update transmission can only take place at the start of an epoch. Without loss of generality, we fix a unit time span for such slots, and focus on a finite time-horizon of $N$ epochs, meant to be sufficiently large to provide a sensible model of operation for real-world sensors.
To track the freshness of information, we define the instantaneous AoI for the source at time $t$ as \cite{yates2021age} 
$  A(t) := t - \sigma(t)$,
where $\sigma(t)$ is the most recent epoch over which an update was received by the sink as of time $t$, and consider the \emph{average AoI} over the time span of interest, i.e.
\begin{equation}
\Delta :=  \frac{1}{N} \int_{0}^N A(t) \,dt.
\label{eq:aoi}
\end{equation}
A possible timeline for the AoI evolution is reported in Fig.~\ref{fig:timeline_aoi}, where one may see that $\Delta$ can be computed as the area below $A(t)$ within the interval $[0,N]$, normalized to the time-horizon.

We assume that, over the considered horizon of operation, the sensor can send a maximum number of updates equal to $M \ll N$. Such a constraint can be due to different reasons, but it is convenient to connect it to limited energy available at the transmitter's side, especially if this is a remote battery-powered sensor placed in a location that is inaccessible (thus making a battery replacement impossible or costly), or an energy-harvesting device, in which case $M$ can be connected to the average energy amount acquired over a recharging cycle.

Finally, update transmissions can be successful or not, depending on the channel state. We assume transmission outcomes to be independent of each other, so that the success of the $i$-th update follows a Bernoulli process with probability $p_i$. For analytical tractability, we will assume a time-invariant distribution, so that $p_i = p$ for all $i = 1,\dots,M$. We note that extensions to different choices of $p_i$'s due to channel correlation or more advanced retransmission techniques are also possible, e.g., considering approaches such as \cite{ceran2019average,B}.

\subsection{Optimization without feedback from the receiver}
We start by studying how to minimize $\Delta$ in the absence of feedback from the receiver. In this case, the goal for the sender is to determine a priori (i.e., offline) how to place the transmission instants in the optimal way. To this aim, if the $M$ update epochs are chosen to be $\tau_1, \tau_2, \dots, \tau_M$, we can define $M{+}1$ intervals $\{ y_i \}$, with $i=0, \dots, M$ such that \mbox{$y_i = \tau_{i{+}1} {-} \tau_i$}, with $\tau_0 {=} 0$ and $\tau_{M{+}1}{=}N$ for consistency.

\begin{figure*}[!t]
\hspace{-0.3cm}
    \subfloat[$p=1$]{
    \includegraphics[width=\figwa]{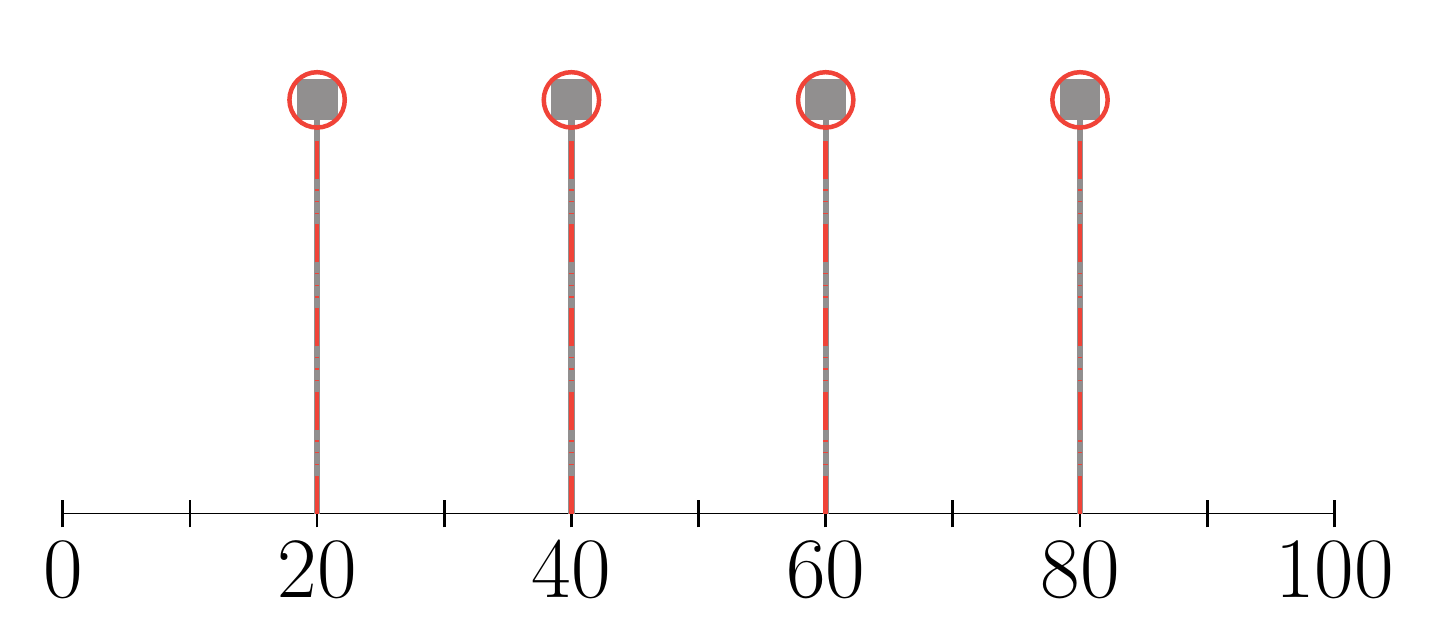}
    \label{fig:plrDiffDistriba1}
    }
    \subfloat[$p=0.8$]{
    \includegraphics[width=\figwa]{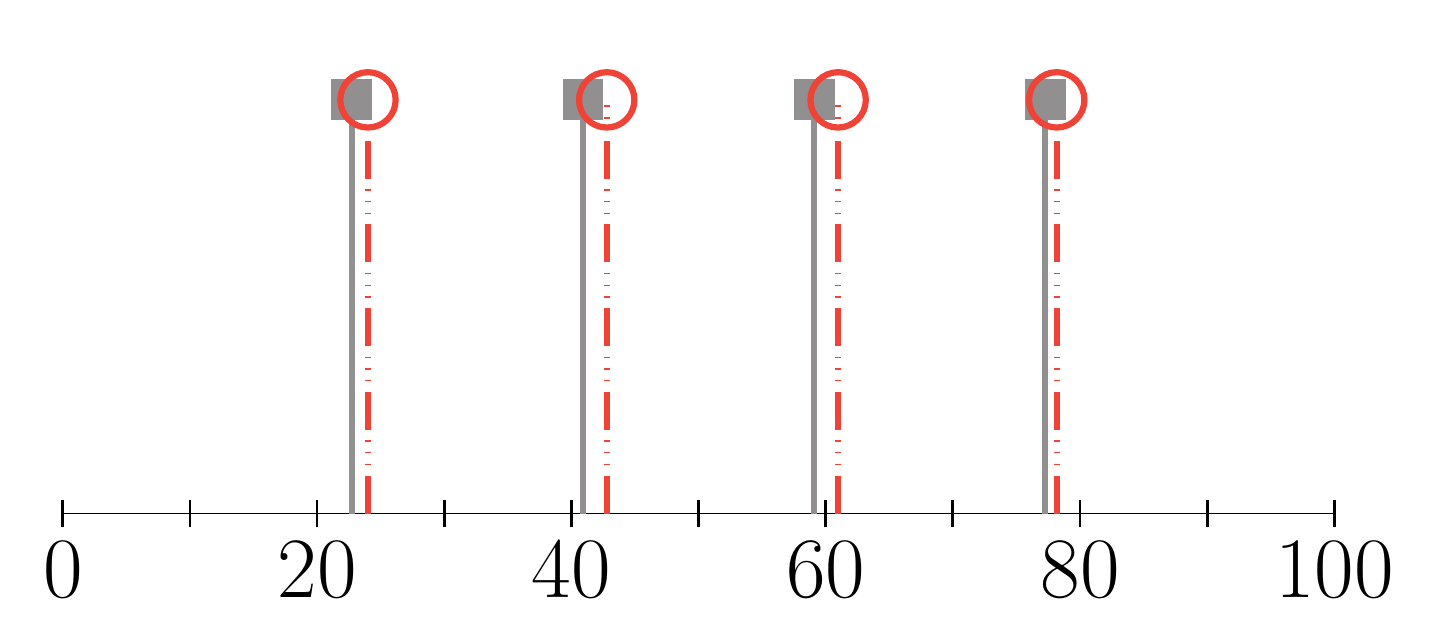}
    \label{fig:plrDiffDistribb1}
    \includegraphics[width=\figwa]{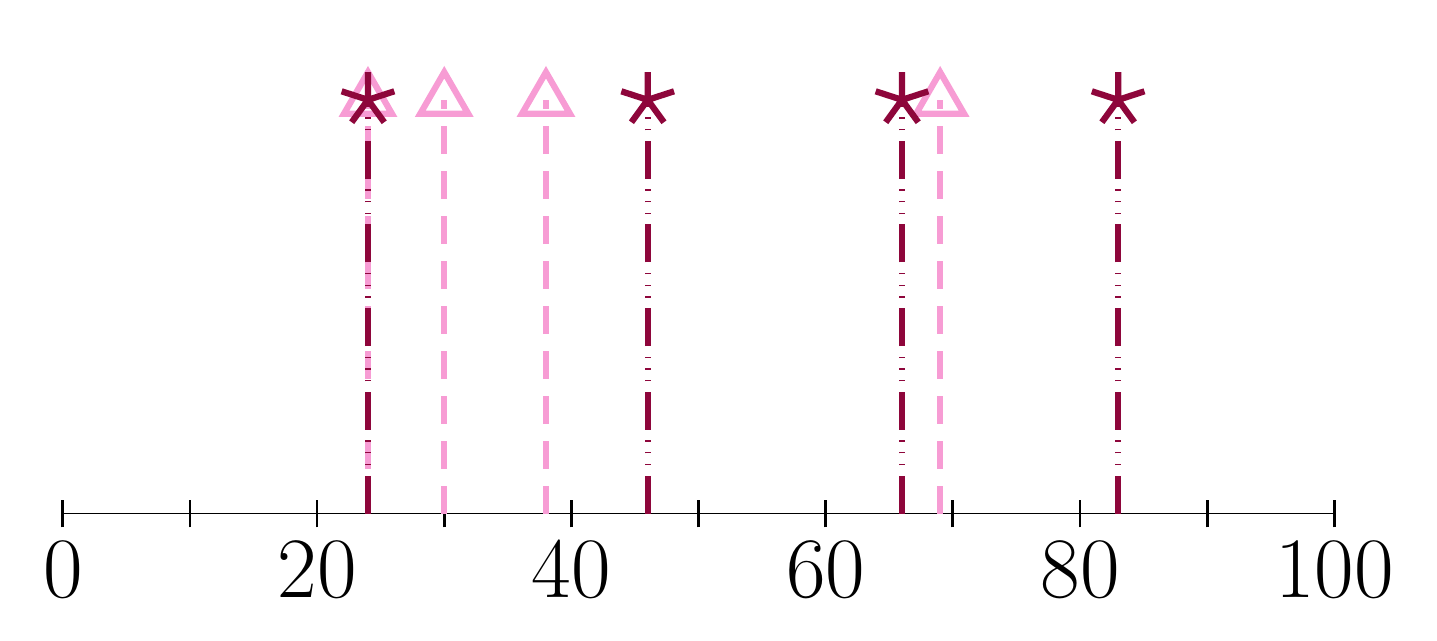}
    \label{fig:truDiffDistribb2}
    }
    \subfloat[$p=0.4$]{
    \includegraphics[width=\figwa]{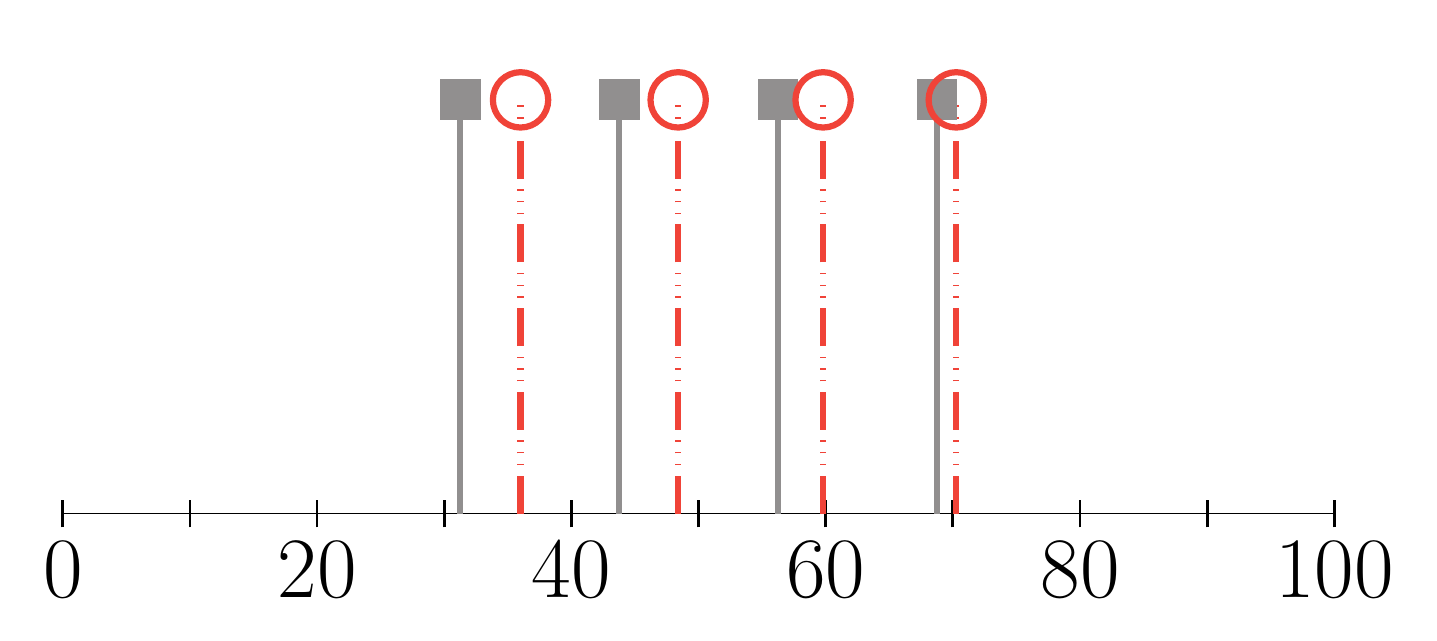}
    \label{fig:truDiffDistribc1}
    }
    \caption{Schedules for $N{=}100$, $m{=}4$, for different values of the success probability $p$. Black stems represent the offline optimization (without feedback). Red dashed stems are the average instants of the scheduled updates in the online optimization with feedback. In (b), the additional plot to the right reports two instances of online schedule based on the outcomes, for $4$ consecutive successful transmissions (dark red,  $*$) or $2$ failures followed by $2$ successes (pink, $\triangle$).}
    \label{fig:plrAndTru}
\end{figure*}

Leaning on this notation, the problem can be restated in terms of the $M+1$ optimization variables $\mathbf{y} = \{ y_i \}$ as
\begin{IEEEeqnarray}{rcl}
\displaystyle \min_{\mathbf y} &\; & \Delta( \mathbf{y} ) \quad\\
\textrm{s.t.} && \sum_{i=0}^{M} y_i = N \nonumber
\end{IEEEeqnarray}

In turn, the average AoI over the finite time horizon can easily be computed as a function of $\mathbf y$ as
\begin{align}
\Delta( \mathbf{y} ) &= \frac{1}{N} \sum_{i=0}^{M} \left[ \frac{{y_i}^2}{2} + \sum_{j=i+1}^{M} y_i y_j (1-p)^{j-i} \right]
\label{eq:deltay}
\end{align}
where the first contribution within brackets accounts for the area of the triangle of side $y_i$ that is always associated to the $i$-th update, whereas the subsequent summation captures the area of the parallelograms of sides $y_i$ and $y_j$ that contribute to the AoI only in case of failure of the $i$-th (and possibly of the following) transmission(s), weighted by the corresponding probabilities.
An example of this reasoning is visually available in Fig.~\ref{fig:timeline_aoi}, where the loss of the second and third updates adds the dashed parallelogram areas $y_1\,y_2$ and $y_1\,y_3$ to the overall computation.

Finding the AoI minimum requires nulling the gradient $\nabla \Delta( \mathbf{y} )$, which, in the specific coordinates $y_i$-s, corresponds to setting the first-order partial derivatives of \eqref{eq:deltay} to $0$, i.e.,
\begin{equation}
\frac{ {\deriva} \Delta( \mathbf{y} )}{ {\deriva} y_i} = 0  \qquad \forall i.
\label{eq:derivative}
\end{equation}
Observing that, by definition, $y_M = N - \sum\nolimits_{i=0}^{M-1} y_i$, \eqref{eq:derivative} leads after simple manipulations to
\begin{IEEEeqnarray}{l}
\Big(2 y_i - N + \sum_{j \neq i} y_j \Big) \big[ 1 - (1{-}p)^{M{-}i} \big] \\
+ \sum_{j \neq i} y_j \Big[ (1{-}p)^{|i{-}j|} - (1{-}p)^{M{-}j} \Big] = 0 \nonumber
\end{IEEEeqnarray}
 for all $i = 0, \dots, M$. We thus obtain a full-rank system of $M$ linear equations in $M$ unknowns, whose solution (easily obtained with standard tools) offers the sought optimal transmission times in the absence of feedback.

\subsection{Optimization when feedback is available}
\label{sub:fed}

Most of the scheduling optimizations for status updates do not perform the previously discussed offline optimization, but rather consider a stateful online procedure where the AoI after each update is known at the transmitter side. This approach would require an instantaneous and cost-free feedback, available at the sensor when scheduling the updates.
In practical systems, however, waiting and receiving a feedback (even if we assume the round-trip time to be so short that it does not cause any lag in the scheduling) has a cost \cite{feng2021age}. In fact, the retrieval of $M$ return messages typically entails an energy expenditure for the sensor that can be of the same order of magnitude than that for transmitting the original packets.

To capture and characterize these aspects, we assume the sink to be able to send such a feedback, received without errors at the source. Incidentally, we note that this is a common assumption in the literature, as the feedback packets are usually limited in size and possibly sent over a different channel. Nonetheless, feedback errors could easily be accounted for by adapting the success probability $p$, see, e.g., \cite{B2}.

We account for the cost of feedback by considering a different value $M_{\rm f}$ of available transmission opportunities over the $N$ epochs, with $M_{\rm f} \leq M$. Specifically, we introduce the \emph{feedback cost coefficient} $\eta > 0$, so that $M_{\rm f} = M / (1{+}\eta)$. In other words, for zero feedback cost, the transmitter is able to use $M_{\rm f} = M$ transmission opportunities, the same for the case without feedback. For a larger $\eta$, $M_{\rm f}$ is decreased accordingly; for example, if receiving a feedback message has the same cost as transmitting an update, i.e., $\eta = 1$, then $M_{\rm f} = M/2$.

Even though the number of transmission opportunities may decrease, the presence of feedback can still offer an advantage to the optimization, in that we can achieve a more efficient schedule of the status updates, following a dynamic programming approach.
This problem can be classically cast on defining a state, a control vector, and a noise component \cite{bertsekas}.
The state of the system at time $t$ is $x(t) = \big( A(t), m(t) \big)$, where $A(t)$ is the instantaneous AoI at time $t$. This is initialized as $A(0) = 0$, whereas 
$m(t)$ is the number of transmission opportunities still available at $t$, for which $m(0){=}M_{\rm f}$. The control of the system $u(t)$ simply results in a binary choice on whether to transmit at epoch $t$, while the noise component is completely captured by the success probability $p$.
With these conventions, the system evolves from $x(t)$ as
\begin{itemize}
\item
$x(t+1) = \big( A(t){+}1, m(t) \big)$ if no update is attempted at time $t$ (i.e., $u(t){=} 0$). When no update is sent at $t$, in fact, the AoI increases by one epoch, and the same number of transmissions remain available to the sensor
\item
$x(t+1) = \big( A(t){+}1, m(t){-}1 \big)$ if $m(t) > 0$ and the sensor sends an update at time $t$ (i.e., $u(t){=}1$), but it is unsuccessful, which happens with probability $1-p$
\item
$x(t+1) = \big( 0, m(t){-}1 \big)$ if $m(t) > 0$ and the sensor sends instead a successful update at time $t$ that resets the AoI, which happens with probability $p$.
\end{itemize}

The problem under study can then be solved by finding the optimal control policy $\mu_t(x(t),p)$ to apply at any state \mbox{$x(t) = \big( A(t), m(t) \big)$}, i.e. the strategy that minimizes the expectation over the time horizon of $N$ epochs of a cost \mbox{$g_t \big( x(t), u(t), p \big) = A(t)$}.

This can be achieved by exploiting Bellman's optimality condition \cite{bertsekas}, since it clearly holds that, if the optimal policy is described by $\mu_0, \mu_1, \dots, \mu_{N{-}1}$,
then for any value of an intermediate state $x(t)$ at time $t$, $0< t < N$, occurring with positive probability, the minimizing policy for the residual cost from $t$ till $N$ is $\mu_t, \dots, \mu_{N{-}1}$.

Specifically, at a given state $x(t)$, the optimal policy is
\begin{align}
\mu_t &\left(x(t) ,p \right) = \mathds{1} \bigg[ (1{-}p) \, R_{t{+}1}\! \left( A(t){+}1, m(t){-}1 \right) \\
&+ p \, R_{t{+}1} \left( 0, m(t){-}1 \right) < R_{t{+}1} \left( A(t){+}1, m(t) \right) \bigg] \nonumber
\end{align}
where
\begin{equation}
R_{t} \big( x(t) \big) = \sum_{i=t}^{N} g_i \Big( x(i), \mu\big(x(i)\big) , p \Big)
\end{equation}
and $\mathds{1}[\cdot]$ is an indicator function, equal to $1$ if the condition inside is true, $0$ otherwise.
In other words, the optimal control at time $t$ is achieved by making the decision that minimizes an expected total cost  equal to the AoI, assuming future decisions are optimally made and averaging over channel errors.

Remarkably, the only actions for the border cases $x(N{-}1) = ( A, m )$ with $m>0$ and $x(t) = \big( A, 0 \big)$ are to transmit and not to transmit, respectively, so one can start by defining $\mu$ for these cases and proceed backwards to find the optimal online scheduling for all reachable states at every $t$.

\subsection{Practical consequences}
Fig.~\ref{fig:plrAndTru} shows an instance of the previous analytical results.
In all cases, $M{=}M_\mathsf{f}{=}4$ updates can be scheduled over a time horizon of $N{=}100$ epochs. In the case of a perfectly reliable channel, Fig.~\ref{fig:plrDiffDistriba1} shows that the optimal update instants are uniformly spread over the time horizon \cite{feng2021age}, and there is no difference between a stateful or stateless optimization, since there is anyway no need for feedback.

In Fig.~\ref{fig:plrDiffDistribb1}, the probability of success is set to $p{=}0.8$ and a difference appears between the offline or online scheduling. In particular, the scheduling instants in the offline optimization shift towards the center of the window of interest. For what concerns the case with feedback, the figure shows the average positions of the scheduling instants, since they clearly depend on the specific realization of the channel.
The average position of the updates across the time window is slightly postponed for the online optimization, since the availability of the feedback can be better exploited and this allows to intervene even at a later stage.
The same trend is confirmed in Fig.~\ref{fig:truDiffDistribc1} for an even lower success probability $p=0.4$.

For $p{=}0.8$, an extra plot in Fig.~\ref{fig:truDiffDistribb2} shows also a comparison of practical realizations of the schedule in the presence of feedback, depending on whether the updates succeed or fail. In more detail, we compare a case where all updates are successful (star markers) with one where the first two transmissions fail (triangle markers).
As visible from the plot, in the latter case, subsequent updates are scheduled much earlier to counteract the missing updates due to the channel failures.

\section{Performance Evaluation}
\label{sec:res}

To gauge the role of feedback in the setting under study, we report and discuss some key trends of interest. Unless otherwise specified, all numerical results have been obtained considering a time horizon of $N=1000$ epochs, and assuming $M=10$ transmissions available to the device when operating without resorting to feedback. Such a configuration is inspired by practical IoT systems such as LoRaWAN, where duty cycles in the order of $1\%$ are typical in the ISM band \cite{lora}.

\begin{figure}
  \centering
  \includegraphics[width=\figw]{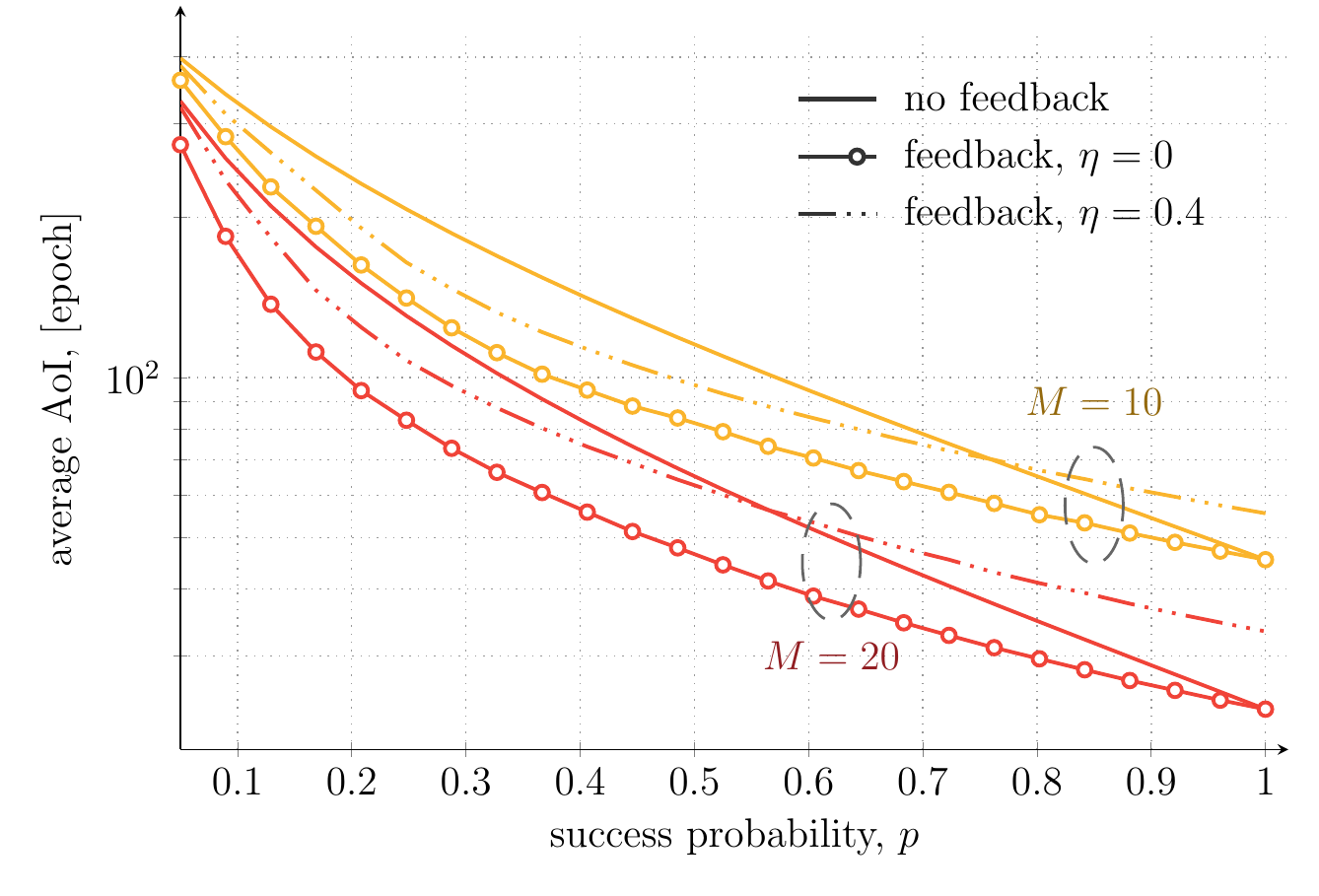}
  \caption{Optimal average AoI vs.\ success probability for different values of $M$ in the absence of feedback (solid lines), with cost-free feedback (circle-marked) and feedback with $\eta{=}0.4$  (dash-dotted). In all cases, $N{=}1000.\!\!$}
  \label{fig:aoiVsPsucc}
\end{figure}

We start by considering Fig.~\ref{fig:aoiVsPsucc}, which shows the average AoI achieved by optimizing the transmission times over the time horizon. Plain solid lines report the behavior in the absence of feedback, whereas circle-marked ones indicate the performance with a cost-free feedback (i.e., $\eta=0$). Finally, dash-dotted lines refer to the use of a costly feedback ($\eta=0.4$), resulting in a reduction of the number of available transmissions from the device to the sink.
In the no-feedback case, the average AoI was directly computed through (\ref{eq:deltay}). Conversely, results in the presence of feedback were obtained applying the dynamic optimization process described in Section \ref{sub:fed} and verified by means of  Montecarlo simulations.

First, consider the case $M{=}10$ (yellow curves), and focus on operations without feedback. As expected, AoI decreases as the success probability increases, thanks to the more frequent delivery of status updates, reaching the minimum value $N/[2(M+1)]$ when $p=1$. In turn, when feedback is available at no cost ($\eta=0$), an improvement emerges. In this case, the possibility to adapt the upcoming transmission times based on the outcome of the current attempt is beneficial, enabling a reduction of the AoI of up to $35\%$. We note that the achievable gain is larger for moderately low success probability values, whereas the two policies behave similarly (and eventually coincide) when $p$ is either very high or very low.

Interestingly, things change significantly when the cost entailed by feedback is accounted for. In fact, while for lower success rates the use of feedback continues to be beneficial, there exist values of $p$ (e.g., $p > 0.75$ in the considered example) for which an offline optimization of the transmission times emerges as the policy of choice. In such conditions, the availability of fewer delivery attempts \--- induced by employing feedback procedures \--- more than counterbalances the positive effects of dynamically adapting the transmission times, rendering the no-feedback approach more effective in terms of AoI. The result offers a first important and non-trivial insight, pinpointing how the use of a return channel shall be carefully considered in practical IoT systems.

Similar trends also emerge when the device is allowed to access the channel more often (case $M=20$, red lines in the plot). In this case, lower values of AoI are attained by all the considered strategies, as a consequence of the more frequent transmissions. The same rationale explains why the use of a costly feedback starts performing worse than the no-feedback counterpart already for lower success probability $p$, thanks to the increased robustness to failures of the latter.

\begin{figure}
  \centering
  \includegraphics[width=\figw]{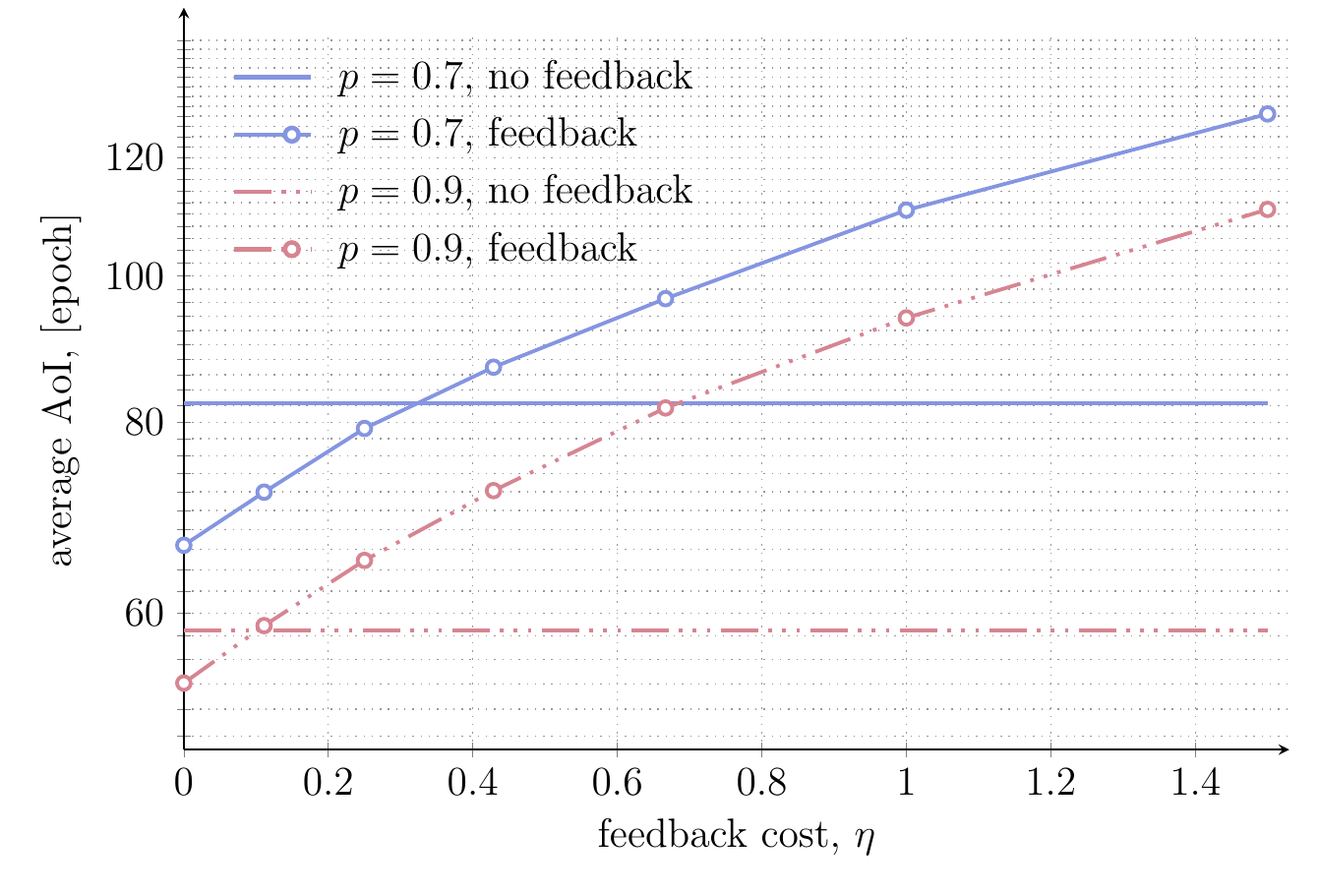}
  \caption{Average AoI vs feedback cost $\eta$ for different success probability $p$. Circle-marked consider stateful optimization (with feedback), non-marked are for no feedback. In all cases, $N{=}1000$, $M{=}10$.}
  \label{fig:aoiVsEta}
\end{figure}

The impact of the feedback cost is further explored in Fig.~\ref{fig:aoiVsEta}, where the average AoI is reported against $\eta$. In the plot, non-marked lines denote the performance of the no-feedback solution (not affected by $\eta$), whereas circle-marked ones refer to the use of feedback. Different results are obtained for $p=0.9$ (dash-dotted red lines) and $p=0.7$ (solid blue lines). The trends highlight how the use of a costly feedback quickly becomes detrimental for the practical values of success probability reported in the figure. From this standpoint, for instance, worse AoI performance is attained already for $\eta = 0.1$ when $p=0.9$. Furthermore, AoI quickly deteriorates with the feedback cost. Interestingly, for $\eta=1$ \--- corresponding to a practically relevant condition in which the reception of feedback may entail an energy cost similar to the one of a transmission \--- a performance loss of up to $60\%$ is experienced for $p=0.9$ in comparison to the simpler no-feedback approach.

\begin{figure}
  \centering
  \vspace{1mm}
  \includegraphics[width=\figw]{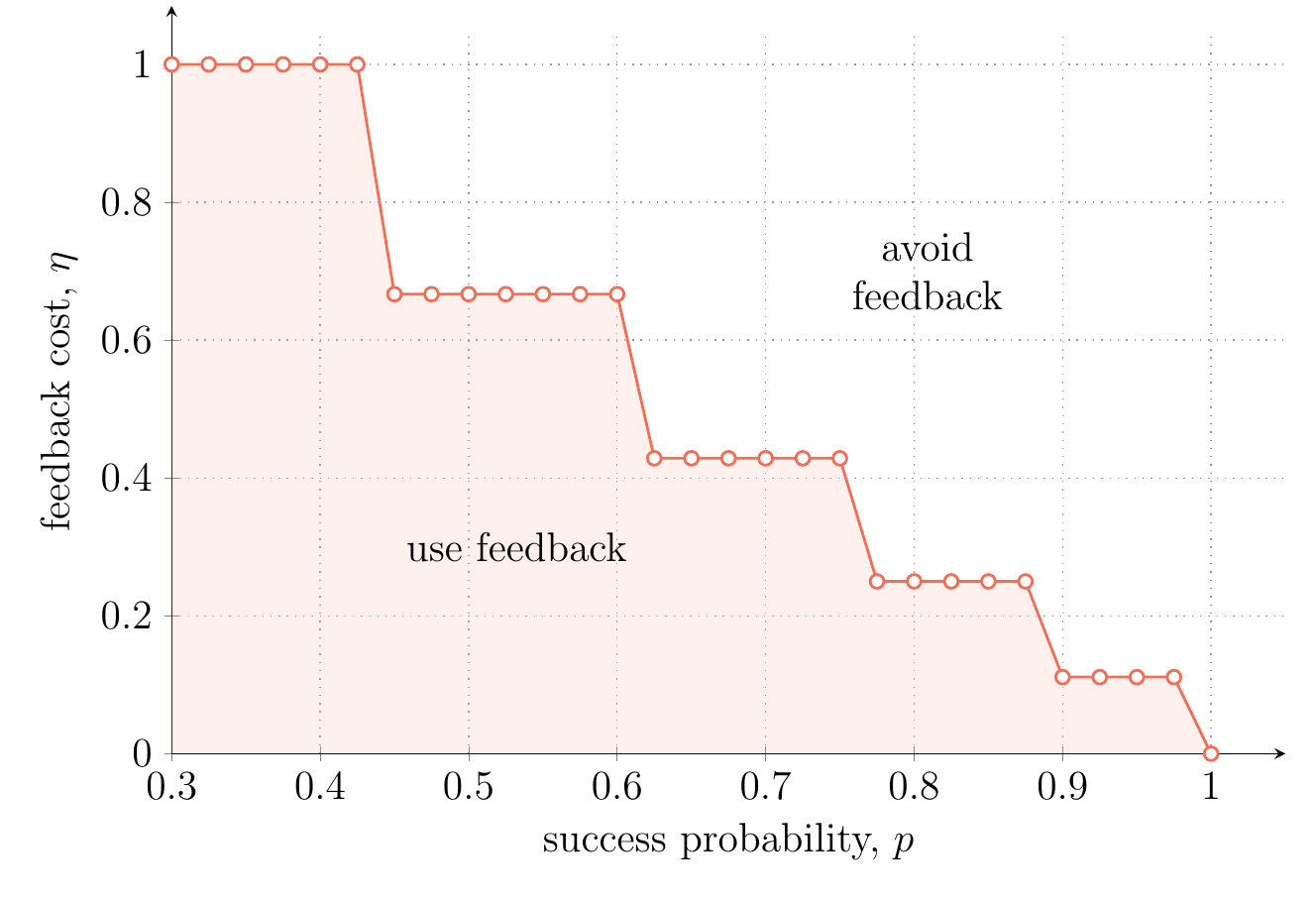}
  \vspace{-1mm}
  \caption{($p_s,\eta$) pairs for which the use of feedback is beneficial or not in terms of attainable average AoI. In all cases, $N=1000$, $M=10$.}
  \label{fig:feedback_region}
\end{figure}

The discussed trends are summarized in Fig.~\ref{fig:feedback_region}, which identifies in the $(p,\eta)$ parameter plane the region where the use of feedback shall or shall not be used from an AoI standpoint. The plot offers a simple yet useful system design tool, quickly identifying the most suitable strategy to be followed under any operating condition. The importance of carefully considering the cost of implementing feedback clearly emerges. In particular, the implementation of a return channel leads to an AoI reduction under harsh channel conditions (i.e., low probability of success), even at the expense of the availability of fewer update delivery attempts. Conversely, a simpler offline optimization is to be preferred when more reliable transmissions can be performed.

Incidentally,  the staircase shape of the plot stems from the fact that a change in performance is only observed when the cost leads to a reduction of the number of available transmissions. In this sense, increasing the time horizon and the number of available transmissions would naturally smooth the curve, without altering the fundamental reported trends.

\section{Conclusions and Future Work}
\label{sec:concs}
We investigated the role of feedback in AoI-optimal finite-horizon scheduling for IoT devices. We developed two different analytical frameworks for stateless and stateful optimization, and we compared them to see whether a feedback from the sink node is required if it comes at a cost (captured by a reduction in the number of available sensor transmissions).
We obtained a comparison of the two scheduling approaches depending on the reliability of the channel, the time horizon, the allowed transmission opportunities, and the feedback cost.

A possible direction to extend the present work would be to expand the time horizon, still keeping it finite, but also allowing for concatenation of different scheduling cycles, which can be a realistic way to represent operation of IoT nodes. The specific comparison between stateless and stateful optimization, and its implications on the utility of exploiting a feedback at the transmitter's side can be extended to more general scenarios also including multiple nodes \cite{zhou2019minimum}, access protocol aspects \cite{sun2019closed}, or energy harvesting \cite{hatami2021aoi}.

\bibliographystyle{IEEEtran}
\bibliography{IEEEabrv,AoI}

\end{document}